\documentclass[9pt]{article}

\usepackage{amsmath,amssymb} 
\usepackage[]{graphicx}
\usepackage{epstopdf}
\usepackage{multirow}
\usepackage{color}
\usepackage{lineno}
\usepackage{ulem}
\usepackage{upgreek}
\usepackage{fullpage}

\def\##1{{\bf #1}}
\def\=#1{\underline{\underline #1}}

\def\eps{\varepsilon}
\def\muo{\mu_{\scriptscriptstyle 0}}
\def\epso{\eps_{\scriptscriptstyle 0}}
\def\etao{\eta_{\scriptscriptstyle 0}}
\def\ietao{\etao^{-1}}
\def\lambdao{\lambda_{\scriptscriptstyle 0}}
\def\ko{k_{\scriptscriptstyle 0}}
\def\co{c_{\scriptscriptstyle 0}}

\def\gammaTI{\gamma_{\rm TI}}
\def\epsTI{\eps_{\rm TI}}
\def\epssubs{\eps_{\rm subs}}
\def\epsCTF{\=\eps_{\rm CTF}}

\def\epsa{\eps_{\rm a}}
\def\epsb{\eps_{\rm b}}
\def\epsc{\eps_{\rm c}}
\def\epsd{\eps_{\rm d}}

\def\LTI{L_{\rm TI}}
\def\LCTF{L_{\rm CTF}}
\def\Lsubs{L_{\rm subs}}
\def\LSigma{L_{\Sigma}}
\def\Lsigma{L_{\sigma}}

\def\ux{\#u_{\rm x}}
\def\uy{\#u_{\rm y}}
\def\uz{\#u_{\rm z}}

\def\sps{\sin\psi}
\def\cps{\cos\psi}
\def\sth{\sin\theta}
\def\cth{\cos\theta}

\newcommand{\SImum}{\ensuremath{\upmu}\textrm{m}}

\newcommand{\rvline}{\hspace*{-\arraycolsep}\vline\hspace*{-\arraycolsep}}
\newcommand{\fraz}{\displaystyle\frac}

\def\tond#1{\left(#1\right)}
\def\quadr#1{\left[#1\right]}
\def\bquadr#1{\Big[#1\Big]}

\def\quote#1{\textquotedblleft #1\textquotedblright}
\newcommand*\diff{\mathop{}\!\mathrm{d}}
\def\largh{0.8\columnwidth}

\def\fref#1{Fig.~\ref{#1}}

\def\sref#1{Sec.~\ref{#1}}

\def\eref#1{eq.~\eqref{#1}}

\def\kinc{\#k_{inc}}
\def\kref{\#k_{ref}}

\def\as{a_{\rm s}}
\def\ap{a_{\rm p}}
\def\rs{r_{\rm s}}
\def\rp{r_{\rm p}}
\def\ts{t_{\rm s}}
\def\tp{t_{\rm p}}

\def\rss{r_{\rm ss}}
\def\rsp{r_{\rm sp}}
\def\rps{r_{\rm ps}}
\def\rpp{r_{\rm pp}}
\def\tss{t_{\rm ss}}
\def\tsp{t_{\rm sp}}
\def\tps{t_{\rm ps}}
\def\tpp{t_{\rm pp}}

\def\Rss{R_{\rm ss}}
\def\Rsp{R_{\rm sp}}
\def\Rps{R_{\rm ps}}
\def\Rpp{R_{\rm pp}}
\def\Tss{T_{\rm ss}}
\def\Tsp{T_{\rm sp}}
\def\Tps{T_{\rm ps}}
\def\Tpp{T_{\rm pp}}

\def\Rab{R_{\rm ab}}
\def\Tab{T_{\rm ab}}

\begin{document}

\begin{center}

\textbf{Enhanced left/right asymmetry in reflection and transmission due to a periodic multilayer of a topological insulator and an anisotropic dielectric material}\\

  \textit{Francesco Chiadini}\\
 
 {Department of Industrial Engineering,
	University of Salerno, via Giovanni Paolo II, 132 -- Fisciano (SA), 
	I-84084, Italy}\\
 
  \textit{Vincenzo Fiumara}\\

{School of Engineering, University of Basilicata, Viale dell'Ateneo Lucano 10, 85100 Potenza, Italy}\\

 \textit{Akhlesh Lakhtakia}\\
{Department of Engineering Science and Mechanics, Pennsylvania State 
University,
	University Park, PA 16802--6812,
	USA}\\
  \textit{Antonio Scaglione}\\
{Department of Industrial Engineering,
	University of Salerno, via Giovanni Paolo II, 132 -- Fisciano (SA), 
	I-84084, Italy}
	
\end{center}
 
 \begin{abstract}
Very weak left/right asymmetry in reflection and transmission is offered by a layer of a topological insulator on top of a layer of an anisotropic  dielectric material, but it can be enhanced very significantly by using a periodic multilayer of both types of materials. This is an attractive prospect for realizing one-way terahertz devices, because both types of materials can be grown 
using standard physical-vapor-deposition techniques.
\end{abstract}

\section*{Introduction}
A topological insulator (TI)~\cite{Hasan,Yan,Ando}  possesses
topologically protected surface states 
leading to an electromagnetic constitution that
must be characterized not only in volumetric terms but also
in terms of a surface admittance~\cite{Mackay}. 
Interest in TIs has greatly grown during the past decade 
as many  materials, such as Bi\textsubscript{2}Se\textsubscript{3} and Sb\textsubscript{2}Te\textsubscript{3}, have been experimentally confirmed to be topological insulators~\cite{Konig,LaForge,CuiZu}.
Mixed materials and new material compositions \cite{Yan,Ando,Barkeshli,DiPietro}  carry promise, especially because the surface admittance can enhanced by the application of a magnetostatic field~\cite{CuiZu,Maciejko}. 

The topologically protected surface states have macroscopic
consequences in optics \cite{Chang,Liu,Liu2,Qi}. Optical modeling of a TI
can be accomplished in two different, though equivalent, ways \cite{Mackay}:
\begin{itemize}
	\item[(i)] as a bi-isotropic  material that is nonreciprocal in the Lorentz sense \cite{Lorentz,Krowne} with the nonreciprocity quantified by a magnetoelectric pseudoscalar denoted by $\gammaTI$~\cite{Liu,Liu2}, or
	\item[(ii)]   as an isotropic dielectric   material with a surface admittance denoted
	by $\gammaTI$~\cite{Mackay}. 
\end{itemize}
From a macroscopic point of view, topological insulation is a phenomenon manifesting itself at the surface but not in the bulk; hence we preferred to model TIs using the surface admittance. This choice also satisfies the Post constraint \cite{Post} that is mandated by the mathematical structure of modern electromagnetic theory.

When light is incident on an infinitely extended layer of a homogeneous material, some is reflected and some is transmitted \cite{Liu,Liu2}. The direction of propagation of the incident light is described by two angles: (a) $\theta\in[0^\circ,90^\circ)$
between the direction of propagation and the normal to the illuminated face of the layer, and (b) $\psi\in[0^\circ,360^\circ)$ between the projection of the direction of propagation on the illuminated
face and a straight line drawn on the face. As a TI is an isotropic dielectric material, the reflectances and transmittances of  a TI layer do not depend on $\psi$ \cite{TI4}. However, if a TI were an anisotropic dielectric material, the reflectances and transmittances would exhibit asymmetry with respect to the reversal of projection of the  
direction of propagation of the incident plane wave on the illuminated face. In other words, if $\theta$ were kept fixed by $\psi$ were to be replaced by $\psi+180^\circ$, the reflectances and transmittances would change~\cite{TI4}.

Theory shows that left/right reflection asymmetry can  be exhibited by a cascade of  a layer of an anisotropic dielectric material and a TI layer \cite{TI+RCWA}. If strong enough,  left/right reflection asymmetry could enable one-way optical devices that can reduce back-scattering noise as well as instabilities in optical communication networks; help efficiently deliver internet at ultrahigh baud rates through lighting fixtures; and sharpen 2D and 3D images for microscopy, tomography, process control, and surgeries. But, in all studies reported thus far,  the magnitude of the surface admittance required
is much greater than the value   that can be effectively achieved~\cite{ICEAA}.

In a bid to enhance left/right reflection asymmetry, we theoretically investigated reflection and transmission characteristics of a periodic multilayer in which identical columnar thin films (CTFs) are interspersed with identical TI layers. A CTF is an ensemble of parallel nanowires aligned obliquely  on a planar substrate~\cite{STFbook,HWbook}. Usually grown by physical vapor deposition \cite{Mattox,Macleod,Raul,Baumeister}, a CTF is a macroscopically homogeneous and  orthorhombic biaxial dielectric material~\cite{Hodg}. The optical response characteristics of CTFs have been exploited for various optical applications~\cite{HWbook}. 

In this paper we report the results of our investigation in which we kept $\gammaTI$ at a low (feasible)
value and we used data for a CTF of tantalum oxide \cite{Hodg,ChMOTL04,ChOC04}.
The paper is organized as follows: in \sref{sec:matmet} we describe in details the materials used and the method to calculate the transmittance/reflectance. In \sref{sec:RaD} results showing the left/right asymmetry in the periodic CTF/TI  multilayer are presented and discussed. Conclusions follow in \sref{sec:cr}.

An $\exp\tond{-i\omega t}$  dependence on time $t$ is implicit, with $\omega$ denoting the angular frequency and $i=\sqrt{-1}$. The free-space wavenumber, the free-space wavelength, and the intrinsic impedance of free space are denoted by $\ko=\omega\sqrt{\epso \muo}$, $\lambdao=2\pi/\ko$, and $\etao=\sqrt{\muo/\epso}$, respectively, with $\epso$ and $\muo$ being the permeability and permittivity of free space. The speed of light in vacuum is denoted by $\co=1/\sqrt{\epso\muo}$, the reduced Planck constant   by $\hbar$, and the charge of an electron by $q_e$.  Vectors are in boldface;  Cartesian unit vectors are identified as $\ux$, $\uy$, and $\uz$;  $\#r=x\ux+y\uy+z\uz$ is the position vector; dyadics are underlined twice; column vectors are in boldface and enclosed in square brackets; and matrices are double underlined and enclosed in square brackets.

\section{Theory}\label{sec:matmet}
We suppose that the half spaces $z<0$ and $z>\LSigma=N\Lambda+\Lsubs$ are occupied by air. The region 
$0<z<\Lsigma=N\Lambda$ is occupied by a periodic multilayer made of $N$ unit cells. The multilayer is of infinite extent in the $xy$ plane. Each unit cell
of thickness $\Lambda=\LTI+\LCTF$
comprises a TI of thickness $\LTI$ and a CTF of thickness $\LCTF$.  
The relative permittivity scalar of the TI is denoted by $\epsTI$.
The 3$\times$3 relative permittivity matrix of
the CTF is expressed as \cite{STFbook,Hodg,ChMOTL04,ChOC04}
\begin{equation}
\epsCTF=\begin{bmatrix}
\epsb + \tond{\epsa-\epsb}\sin^2\chi&\hspace{15pt} 0\hspace{15pt} & -\fraz{1}{2}\tond{\epsa-\epsb}\sin 2\chi
\\[8pt]
0&\hspace{15pt}  \epsc\hspace{15pt} &0
\\[8pt]
-\fraz{1}{2}\tond{\epsa-\epsb}\sin2\chi&\hspace{15pt}  0\hspace{15pt}  & \epsa - \tond{\epsa-\epsb}\sin^2\chi
\end{bmatrix}\,,
\label{eq:epsCTF}
\end{equation}
where $\eps_{\rm a,b,c}$ are the eigenvalues of $\=\eps_{CTF}$
and the angle $\chi\in(0^\circ,90^\circ]$.
The region $\Lsigma<z<\LSigma$ is occupied by a dielectric material functioning as a substrate,
its relative permittivity scalar being denoted by $\epssubs$.

\begin{figure}[!h]
	\begin{center}
		\includegraphics[width=\largh]{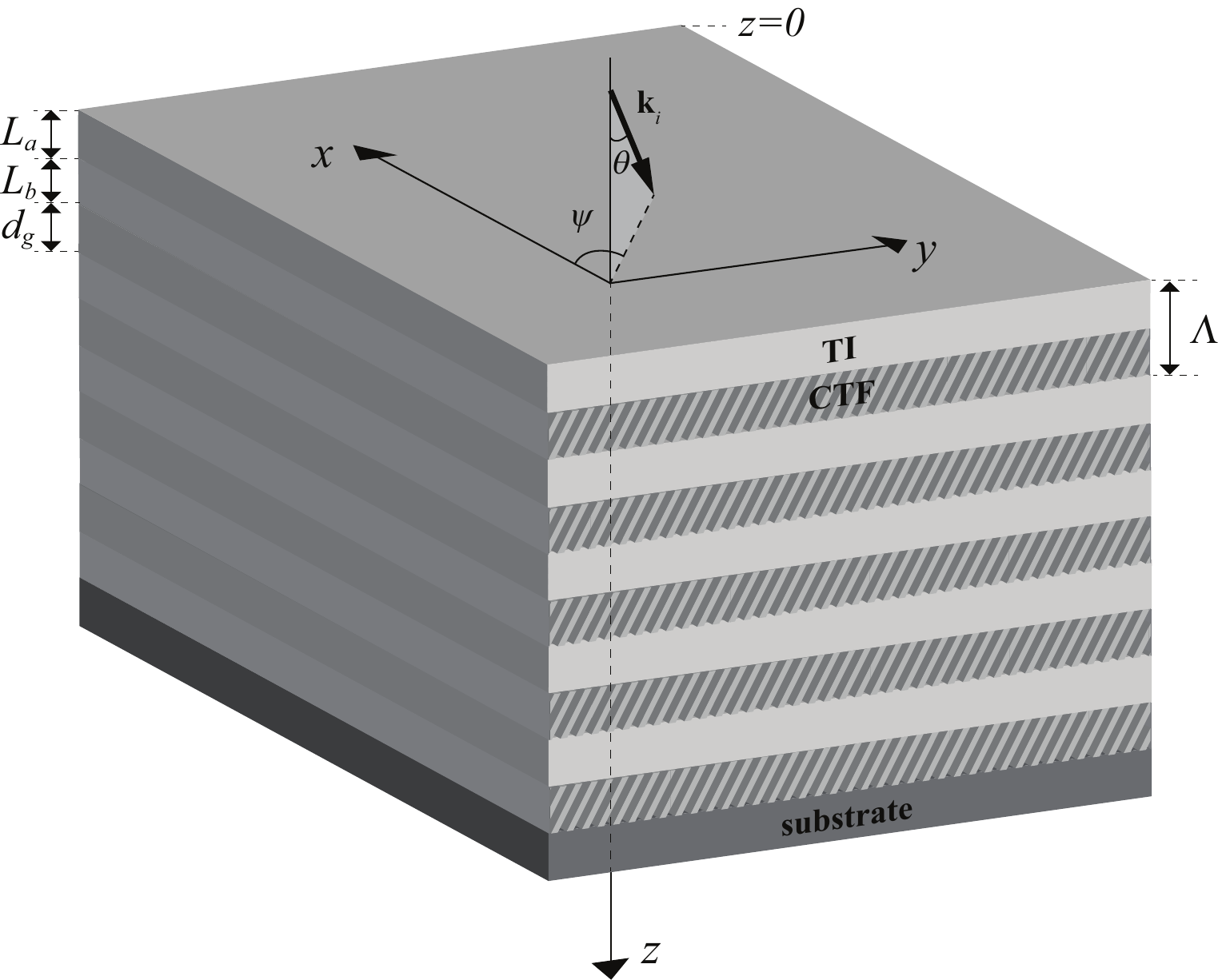}
		\caption{Schematic of the boundary-value problem solved.  The structure shown has
			$N=5$ unit cells and is infinitely extended in the $xy$ plane.}
		\label{fig:schematic}
\end{center}\end{figure}

A plane wave is incident at an angle $\theta$ with respect to the $z$ axis and at an angle $\psi$ with respect to the $x$ axis in the $xy$ plane, as   illustrated in \fref{fig:schematic}.
The wave vector of the incident plane wave can therefore be written as
\begin{equation}
\kinc = \kappa\left(\ux\cps+\uy\sps\right)+\uz\ko\cth\,,
\label{eq:kinc}
\end{equation}
where $\kappa=\ko\sth$. The electric and magnetic field phasors of the incident plane wave are
given by
\begin{equation}
\begin{cases}{}
\#E_{inc}\tond{\#r}&=\#E_{0,inc}\exp\tond{i\kinc\cdot\#r}\\[10pt]
\#H_{inc}\tond{\#r}&=\#H_{0,inc}\exp\tond{i\kinc\cdot\#r}
\end{cases},\quad z<0\,,
\end{equation}
where the amplitude vectors are represented by column 3-vectors as  
\begin{equation}
\quadr{\#E_{0,inc}}=
\begin{bmatrix}
E_{x,inc}\\
E_{y,inc}\\
E_{z,inc}
\end{bmatrix}=
\begin{bmatrix}
-\sps &-\cps\cth\\
\cps& -\sps\cth\\
0 & \sth
\end{bmatrix}
\begin{bmatrix}
\as\\ \\
\ap
\end{bmatrix}
\label{eq:Ei}\end{equation}
and
\begin{equation}
\quadr{\#H_{0,inc}}=
\begin{bmatrix}
H_{x,inc}\\
H_{y,inc}\\
H_{z,inc}
\end{bmatrix}=
\ietao\begin{bmatrix}
-\cps\cth&\sps\\
-\sps\cth&-\cps\\
\sth& 0
\end{bmatrix}
\begin{bmatrix}
\as\\ \\
\ap
\end{bmatrix}\,.
\label{eq:Hi}
\end{equation}
The scalar coefficients
with $\as$ and $\ap$ of the $s$- and  $p$-polarized components, respectively, are assumed
to be known \cite{STFbook,Chen}.

The wave vector of the reflected plane wave is consequently given by
\begin{equation}
\kref = \kappa\left(\ux\cps +\uy\sps\right)-\uz\ko\cth\,,
\label{eq:kref}
\end{equation}
and the electric and magnetic field phasors as
\begin{equation}
\begin{cases}{}
\#E_{ref}\tond{\#r}&=\#E_{0,ref}\exp\tond{i\kref\cdot\#r}\\[10pt]
\#H_{ref}\tond{\#r}&= \#H_{0,ref}\exp\tond{i\kref\cdot\#r}
\end{cases},\quad z<0\,,
\end{equation}
where  the amplitude vectors
\begin{equation}
\quadr{\#E_{0,ref}}=
\begin{bmatrix}
E_{x,ref}\\
E_{y,ref}\\
E_{z,ref}
\end{bmatrix}=
\begin{bmatrix}
-\sps& \cps\cth\\
\cps& \sps\cth\\
0& \sth
\end{bmatrix}
\begin{bmatrix}
\rs\\ \\
\rp
\end{bmatrix}
\label{eq:Er}
\end{equation}
and
\begin{equation}
\quadr{\#H_{0,ref}}=
\begin{bmatrix}
H_{x,ref}\\
H_{y,ref}\\
H_{z,ref}
\end{bmatrix}=
\ietao\begin{bmatrix}
\cps\cth&\sps\\
\sps\cth&-\cps\\
\sth& 0
\end{bmatrix}
\begin{bmatrix}
\rs\\ \\
\rp
\end{bmatrix}
\label{eq:Hr}\end{equation}
employ $\rs$ and $\rp$ as the unknown coefficients of the $s$- and  $p$-polarized components, respectively \cite{STFbook,Chen}.

The wave vector of the transmitted plane wave is exactly the same as that of the incident plane
wave. Therefore, the
electric and magnetic field phasors of the transmitted plane wave are
\begin{equation}
\begin{cases}{}
\#E_{trs}\tond{\#r}&=\#E_{0,trs}
\exp\left[i \kinc\cdot\left(\#r-\LSigma\uz\right)\right]
\\[10pt]
\#H_{trs}\tond{\#r}&= \#H_{0,trs}
\exp\left[i \kinc\cdot\left(\#r-\LSigma\uz\right)\right]
\end{cases},\quad z>\LSigma\,,
\end{equation}
where the column 3-vectors  
\begin{equation}
\quadr{\#E_{0,trs}}=
\begin{bmatrix}
E_{x,trs}\\
E_{y,trs}\\
E_{z,trs}
\end{bmatrix}=
\begin{bmatrix}
-\sps&-\cps\cth\\
\cps&-\sps\cth\\
0& \sth
\end{bmatrix}
\begin{bmatrix}
\ts\\ \\
\tp
\end{bmatrix}\,
\label{eq:Et}\end{equation}
and
\begin{equation}
\quadr{\#H_{0,trs}}=
\begin{bmatrix}
H_{x,trs}\\
H_{y,trs}\\
H_{z,trs}
\end{bmatrix}=
\ietao\begin{bmatrix}
-\cps\cth&\sps\\
-\sps\cth&-\cps\\
\sth&0
\end{bmatrix}
\begin{bmatrix}
\ts\\ \\
\tp
\end{bmatrix}\,
\label{eq:Ht}\end{equation}
contain $\ts$ and $\tp$ as the unknown coefficients of the $s$- and   $p$-polarized components, respectively \cite{STFbook,Chen}. A boundary-value problem has to be solved
in order to determine the coefficients   $\rs$, $\rp$, $\ts$, and $\tp$ in terms of $\as$ and $\ap$.

The electric and magnetic field phasors everywhere are conveniently represented as \cite{STFbook}
\begin{equation}
\begin{cases}{}
\#E\tond{\#r}=\#e\tond{z}\exp\quadr{i\kappa\tond{x\cps+y\sps}}\\[10pt]
\#H\tond{\#r}=\#h\tond{z}\exp\quadr{i\kappa\tond{x\cps+y\sps}}
\end{cases}\,. 
\label{eq:EH}\end{equation}
Furthermore, we define
the column 4-vector
\begin{equation}
\quadr{\#f\tond{z}}=\begin{bmatrix}
\ux\cdot\#e\tond{z}\\
\uy\cdot\#e\tond{z}\\
\ux\cdot\#h\tond{z}\\
\uy\cdot\#h\tond{z}
\end{bmatrix}\,.
\label{eq:fz}
\end{equation}

Together, \eref{eq:Ei}, \eref{eq:Hi}, \eref{eq:Er}, and \eref{eq:Hr} yield
\begin{equation}
\bquadr{\#f\tond{0^-}}=\bquadr{\=K}\begin{bmatrix}
\as\\
\ap\\
\rs\\
\rp
\end{bmatrix}\,,
\label{eq:f0}\end{equation}
where the 4$\times$4 matrix
\begin{equation}
\quadr{\=K}=
\begin{bmatrix}
-\sps & \cps\cth & -\sps& \cps\cth\\[5pt]
-\cps & -\sps\cth& \cps& \sps\cth\\[5pt]
-\ietao\cps\cth & \ietao\sps & \ietao\cps\cth& \ietao\sps\\[5pt]
-\ietao\sps\cth & \ietao\cps & \ietao\sps\cth& -\ietao\cps
\end{bmatrix}\,.
\end{equation}
Together, \eref{eq:Et} and \eref{eq:Ht} yield
\begin{equation}
\quadr{\#f\tond{\LSigma^+}}=\quadr{\=K}\begin{bmatrix}
\ts\\
\tp\\
0\\
0
\end{bmatrix}
\label{eq:ft}
\end{equation}

The $n$th unit cell, $n\in[1,N]$, occupies the region $z_{n-1}<z<z_n$, where
$z_n=n\Lambda$. The region $z_{n-1}<z<\zeta_n=z_{n-1}+\LTI$ is occupied by the chosen TI. In this region,
$\quadr{\#f\tond{z}}$ obeys the 4$\times$4 matrix ordinary differential equation
\begin{equation}
\fraz{\diff}{\diff z}\bquadr{\#f\tond{z}}=i\bquadr{\=P}_{TI}\bquadr{\#f\tond{z}}\,,
\quad z_{n-1}<z<\zeta_n\,,
\label{eq:ODE-TI}\end{equation}
where the 4$\times$4 matrix  
\begin{eqnarray}
\nonumber
&&
\bquadr{\=P}_{TI}=\omega\left[\begin{array}{cccc}
0 & 0 & 0 & \muo \\[6pt]
0 & 0 & -\muo & 0 \\[6pt]
0 & - \epso \epsTI & 0 & 0\\[10pt]
\epso\epsTI  & 0 & 0 & 0
\end{array}\right]
\\[5pt]
\nonumber
&& \quad
+\frac{\kappa^2}{\omega\epso\epsTI}\,
\left[\begin{array}{cccc}
0 & 0 & \cos\psi\,\sin\psi & -\cos^2\psi \\[4pt]
0 & 0 & \sin^2\psi & -\cos\psi\,\sin\psi \\[4pt]
0 & 0& 0 & 0\\[4pt]
0 & 0 & 0 & 0
\end{array}\right]
\\[5pt]
\label{PmatchiTI}
&& \quad
+\frac{\kappa^2}{\omega\muo}\,
\left[\begin{array}{cccc}
0 & 0 & 0 & 0\\[4pt]
0 & 0 & 0 & 0\\[4pt]
-\cos\psi\,\sin\psi & \cos^2\psi & 0 & 0\\[4pt]
-\sin^2\psi & \cos\psi\,\sin\psi & 0 & 0
\end{array}\right]\,.
\end{eqnarray}
The solution of \eref{eq:ODE-TI} delivers \cite{Hoch}
\begin{equation}
\bquadr{\#f\tond{\zeta_n^-}}=\bquadr{\=Q}_{TI}\bquadr{\#f\tond{z_{n-1}^+}}\,,
\end{equation}
where the 4$\times$4 matrix
\begin{equation}
\bquadr{\=Q}_{TI} = \exp\left\{
i\bquadr{\=P}_{TI}\LTI
\right\}\,.
\end{equation}

The region $\zeta_n<z<z_n$ is occupied by the chosen CTF. In this region,
$\quadr{\#f\tond{z}}$ obeys the 4$\times$4 matrix ordinary differential equation \cite{STFbook}
\begin{equation}
\fraz{\diff}{\diff z}\bquadr{\#f\tond{z}}=i\bquadr{\=P}_{CTF}\bquadr{\#f\tond{z}}\,,
\quad \zeta_n<z<z_n\,,
\label{eq:ODE-CTF}\end{equation}
where the 4$\times$4 matrix  
\begin{eqnarray}
\nonumber
&&\hspace{-70pt}
\bquadr{\=P}_{CTF}=\omega\left[\begin{array}{cccc}
0 & 0 & 0 & \muo \\[6pt]
0 & 0 & -\muo & 0 \\[6pt]
0 & - \epso \epsc & 0 & 0\\[10pt]
\epso\epsd  & 0 & 0 & 0
\end{array}\right]
\\[5pt]
\nonumber
&& \hspace{-70pt}
+\,\kappa\,\frac{\epsd\left(\epsa-\epsb\right)   }{\epsa\,\epsb}\,
\frac{\sin 2\chi}{2}\,
\left[\begin{array}{cccc}
\cos\psi & 0 & 0 & 0\\[10pt]
\sin\psi &0 & 0 & 0\\[10pt]
0 & 0 & 0& 0\\[10pt]
0 & 0 & - \sin\psi &  \cos\psi
\end{array}\right]
\\[5pt]
\nonumber
&&  \hspace{-70pt}
+\,\frac{\kappa^2}{\omega\epso}\,
\frac{\epsd}{\epsa\,\epsb}\,
\left[\begin{array}{cccc}
0 & 0 & \cos\psi\,\sin\psi & -\cos^2\psi \\[4pt]
0 & 0 & \sin^2\psi & -\cos\psi\,\sin\psi \\[4pt]
0 & 0& 0 & 0\\[4pt]
0 & 0 & 0 & 0
\end{array}\right]
\\[5pt]
\label{PmatchiCTF}
&& \hspace{-70pt}
+\,\frac{\kappa^2}{\omega\muo}\,
\left[\begin{array}{cccc}
0 & 0 & 0 & 0\\[4pt]
0 & 0 & 0 & 0\\[4pt]
-\cos\psi\,\sin\psi & \cos^2\psi & 0 & 0\\[4pt]
-\sin^2\psi & \cos\psi\,\sin\psi & 0 & 0
\end{array}\right]\,
\end{eqnarray}
employs
\begin{equation}
\epsd=\fraz{\epsa\epsb}{\epsa\cos^2\chi+\epsb\sin^2\chi}
\,.
\label{eq:epsd}
\end{equation}
The solution of \eref{eq:ODE-CTF} delivers \cite{Hoch}
\begin{equation}
\bquadr{\#f\tond{z_n^-}}=\bquadr{\=Q}_{CTF}\bquadr{\#f\tond{\zeta_n^+}}\,,
\end{equation}
where the 4$\times$4 matrix
\begin{equation}
\bquadr{\=Q}_{CTF} = \exp\left\{
i\bquadr{\=P}_{CTF}\LCTF
\right\}\,.
\end{equation}
As shown elsewhere~\cite{TI+RCWA}, consideration of the boundary conditions at the
the interface $z=\zeta_n$ of the TI layer and the CTF yields
\begin{equation}
\quadr{\#f\tond{\zeta_n^+}}=\quadr{\=V}\quadr{\#f\tond{\zeta_n^-}}
\label{eq:transf2a}
\end{equation}
where the 4$\times$4 matrix
\begin{equation}
\bquadr{\=V}=
\begin{bmatrix}
1&0&0&0\\
0&1&0&0\\
-\gamma_{TI}&0&1&0\\
0&-\gamma_{TI}&0&1\\
\end{bmatrix}\,.
\label{eq:V}
\end{equation}
Accordingly,
\begin{equation}
\bquadr{\#f\tond{z_n^-}}=
\bquadr{\=Q}_{CTF}\quadr{\=V} 
\bquadr{\=Q}_{TI}\bquadr{\#f\tond{z_{n-1}^+}}\,.
\label{eq:25}
\end{equation}
Consideration of the boundary conditions at the interface $z=z_{n-1}$ yields~\cite{TI+RCWA}
\begin{equation}
\quadr{\#f\tond{z_{n-1}^+}}=\quadr{\=V}^{-1}\quadr{\#f\tond{z_{n-1}^-}}\,.
\label{eq:transf2b}
\end{equation}
From the last two equations, we obtain
\begin{equation}
\bquadr{\#f\tond{z_n^-}}=
\bquadr{\=U} \bquadr{\#f\tond{z_{n-1}^-}}\,,
\label{eq:27}
\end{equation}
where the 4$\times$4 matrix
\begin{equation}
\bquadr{\=U}=
\bquadr{\=Q}_{CTF}\quadr{\=V} 
\bquadr{\=Q}_{TI}\quadr{\=V}^{-1}
\label{def-U}
\end{equation}
is the characteristic matrix of a unit cell.
Application of \eref{eq:27} repeatedly from $n=1$ to $n=N$ delivers
\begin{equation}
\bquadr{\#f\tond{\Lsigma^-}}=\bquadr{\=U}^N \bquadr{\#f\tond{0^-}}\,.
\label{eq:29}
\end{equation}

The region $\Lsigma<z<\LSigma$ is occupied by the substrate. In this region,
$\quadr{\#f\tond{z}}$ obeys the 4$\times$4 matrix ordinary differential equation
\begin{equation}
\fraz{\diff}{\diff z}\bquadr{\#f\tond{z}}=i\bquadr{\=P}_{subs}\bquadr{\#f\tond{z}}\,,
\quad \Lsigma<z<\LSigma\,,
\label{eq:ODE-subs}\end{equation}
where the 4$\times$4 matrix $\bquadr{\=P}_{subs}$ is obtained by replacing
$\epsTI$ by $\epssubs$ on the right side of \eref{PmatchiTI}.
The solution of \eref{eq:ODE-subs} delivers \cite{Hoch}
\begin{equation}
\bquadr{\#f\tond{\LSigma^-}}=\bquadr{\=Q}_{subs}\bquadr{\#f\tond{\Lsigma^+}}\,,
\label{eq:31}
\end{equation}
where the 4$\times$4 matrix
\begin{equation}
\bquadr{\=Q}_{subs} = \exp\left\{
i\bquadr{\=P}_{subs}\Lsubs
\right\}\,.
\end{equation}
Application of the standard boundary conditions \cite{Chen} at the interface $z=\Lsigma$ yields
\begin{equation}
\quadr{\#f\tond{\Lsigma^+}}= \quadr{\#f\tond{\Lsigma^-}}\,,
\label{eq:33}
\end{equation}
so that
\begin{equation}
\bquadr{\#f\tond{\LSigma^-}}=\bquadr{\=Q}_{subs}\bquadr{\=U}^N \bquadr{\#f\tond{0^-}}
\label{eq:34}
\end{equation}
follows from \eref{eq:29} and \eref{eq:31}. Finally,
application of the standard boundary conditions \cite{Chen} at the interface $z=\Lsigma$ yields
\begin{equation}
\quadr{\#f\tond{\LSigma^+}}= \quadr{\#f\tond{\LSigma^-}}\,,
\label{eq:35}
\end{equation}
leading to
\begin{equation}
\bquadr{\#f\tond{\LSigma^+}}=\bquadr{\=Q}_{subs}\bquadr{\=U}^N \bquadr{\#f\tond{0^-}}\,.
\label{eq:36}
\end{equation}

Combining \eref{eq:f0}, \eref{eq:ft}, and \eref{eq:36}, we get
\begin{equation}
\begin{bmatrix}
\ts\\
\tp\\
0\\
0
\end{bmatrix}
=\quadr{\=M} 
\begin{bmatrix}
\as\\
\ap\\
\rs\\
\rp
\end{bmatrix}\,,
\label{eq:37}
\end{equation}
where the 4$\times$4 matrix
\begin{equation}\delimitershortfall=-1pt
\quadr{\=M}=\quadr{\=K}^{-1}\bquadr{\=Q}_{subs}\bquadr{\=U}^N\bquadr{\=K}\,
\label{eq:M}
\end{equation}
can be partitioned into 4 2$\times$2 submatrices as follows:
\begin{equation}
\quadr{\=M}=\begin{bmatrix}
&\quadr{\=M_{11}} & \rvline & \quadr{\=M_{12}}&\\
&& \rvline &&\\
\cline{2-4}
&& \rvline &&\\
&\quadr{\=M_{21}}& \rvline & \quadr{\=M_{22}}  & 
\end{bmatrix} \,.
\label{eq:Mblock}
\end{equation}
Then the scalar coefficients of the reflected plane wave emerge
\begin{equation}
\begin{bmatrix}
\rs\\
\rp
\end{bmatrix}=
-\quadr{\=M_{22}}^{-1}\quadr{\=M_{21}}
\begin{bmatrix}
\as\\
\ap
\end{bmatrix}
\label{eq:Bcoeff}
\end{equation}
and the scalar coefficients of the transmitted plane wave can be calculated as
\begin{equation}
\begin{bmatrix}
\ts\\
\tp
\end{bmatrix}= \quadr{\=M_{tot}}
\begin{bmatrix}
\as\\
\ap
\end{bmatrix}
\label{eq:Ccoeff}
\end{equation}
where 
\begin{equation}
\quadr{\=M_{tot}}= \quadr{\=M_{11}}-\quadr{\=M_{12}}\quadr{\=M_{22}}^{-1}\quadr{\=M_{21}}
\label{eq:Mtot}
\end{equation}
is the whole transfer matrix.

Four reflection coefficients  ${r_{\rm ab}}$ and   four transmission coefficients
${t_{\rm ab}}$, $ a\in\left\{p,s\right\}$ and 
$b\in\left\{p,s\right\}$,   appear in the following relations:
\begin{equation}
\label{r-t}
\left.\begin{array}{ll}
\rs = \rss\,\as+\rsp\,\ap\,,\qquad & \ts = \tss\,\as+\tsp\,\ap
\\[5pt]
\rp = \rps\,\as+\rpp\,\ap\,,\qquad & \tp = \tps\,\as+\tpp\,\ap
\end{array}\right\}\,.
\end{equation}
Accordingly,
four reflectances
are defined as $\Rsp=\vert\rsp\vert^2$, etc., and four transmittances
as $\Tsp=\vert\tsp\vert^2$, etc.
The principle of conservation of
energy requires that
\begin{equation}
\left.\begin{array}{l}
0\leq\Rss+\Rps +\Tss+\Tps \leq 1
\\[5pt]
0\leq\Rpp+\Rsp+\Tpp+\Tsp\leq1
\end{array}
\right\}\,.
\end{equation}
The differences $1-\left(\Rss+\Rps+\Tss+\Tps\right)$ and $1-\left(\Rpp+\Rsp+\Tpp+\Tsp\right)$
indicate the fraction of the incident energy that is absorbed in the region $0<z<\LSigma$.

In order to quantitate left/right reflection asymmetry, we define the functions
\begin{equation}
\Delta{R_{\rm ab}}(\theta,\psi)= {R_{\rm ab}}(\theta,\psi)-{R_{\rm ab}}(\theta,\psi+180^\circ)\,,
\quad
a\in\left\{s,p\right\}\,,\quad
b\in\left\{s,p\right\}\,.
\end{equation}
Likewise,  we define the functions
\begin{equation}
\Delta{T_{\rm ab}}(\theta,\psi)= {T_{\rm ab}}(\theta,\psi)-{T_{\rm ab}}(\theta,\psi+180^\circ)\,,
\quad
a\in\left\{s,p\right\}\,,\quad
b\in\left\{s,p\right\}\,,
\end{equation}
in order to quantitate left/right transmission asymmetry.

\section{Numerical Results}\label{sec:RaD}
Intrinsic TIs are characterized by $\gammaTI=\pm \alpha/\etao$,
where $\alpha =\left(q_e^2/\hbar{\co}\right)/4\pi\epso$ is the (dimensionless) fine structure
constant \cite{Hasan,Yan,Maciejko,Qi}. Either immersion in a magnetostatic field or
a very thin coating of a magnetic material can be used to realize
$\gammaTI=(2m+1)\alpha/\etao$, $m\in\left\{0,\pm1,\pm2,\pm3,\dots\right\}$. 
Thus, the normalized surface admittance $\overline{\gamma} =\gammaTI\etao/\alpha$
can be either a negative or a positive integer. Whereas intrinsic TIs have $\overline{\gamma}=\pm1$,
exploitation of a magnetic field or material 
may increase $\vert\tilde{\gamma}\vert$ realistically to   $2$ or $3$ \cite{Maciejko}. We fixed
$\overline{\gamma}=1$, $\epsTI=3$, and $\LTI=649.5$~nm for all results reported here.

TIs have bandgaps not exceeding 300 meV; hence, we fixed our attention to $\lambdao\in\quadr{4,5}~\SImum$ for calculations. 
We chose the substrate material  to be silicon with relative permittivity $\epssubs=11.68$ and thickness $\Lsubs=5~\SImum$. Furthermore, we set  $\chi=48.50^\circ$,
$\epsa=2.2532$, $\epsb=2.7737$, and $\epsc=2.5475$, based
on data reported for columnar thin films of tantalum oxide~\cite[Sec.~7.3.3]{STFbook}. We also fixed
$\LCTF=749.5$~nm. Thus, $\LTI=\lambdao/4\sqrt{\epsTI}$
and $\LCTF=\lambdao/4\sqrt{\epsa}$  when $\lambdao=4.5~\SImum$.

With $N=1$, the maximum values of $\Delta\Rab$ and $\Delta\Tab$,
${\rm a}\in\left\{s,p\right\}$ and ${\rm b}\in\left\{s,p\right\}$, did not exceed $10^{-2}$ for
any combination of $\theta$ and $\phi$. Although present, such degrees of left/right
asymmetries are not unlikely to be technologically exploitable.

\begin{figure}[h]
	\centering
	\includegraphics[width=\largh]{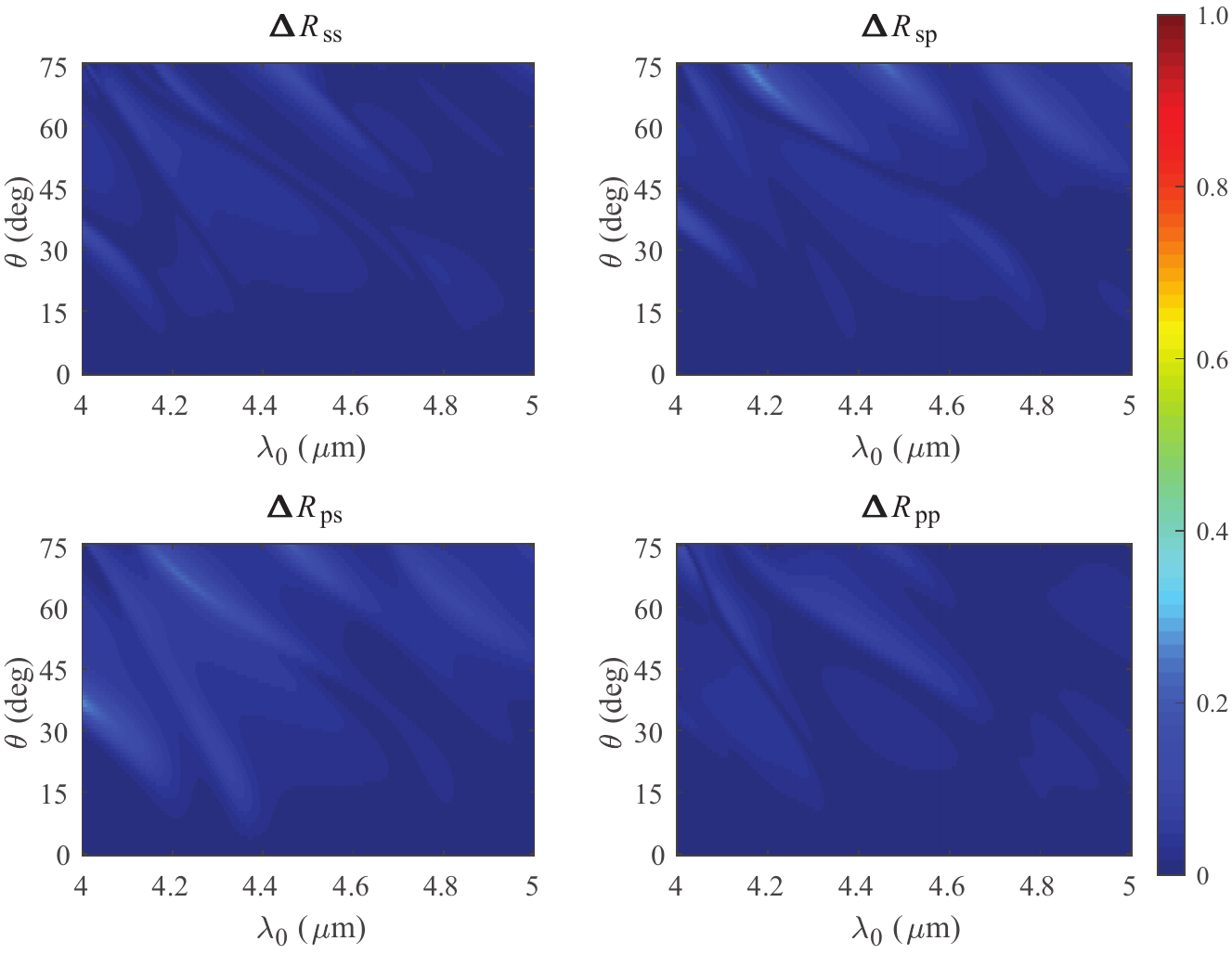}
	\caption{Left/right reflection
		asymmetry functions $\Delta \Rss$, $\Delta \Rsp$, $\Delta \Rpp$, 
		and $\Delta \Rps$ for $\theta\in[0^\circ,90^\circ)$ and $\lambdao\in\quadr{4,5}~\SImum$, when $\psi=45^\circ$ and $N=10$.}
	\label{fig:DeltaR_10}
\end{figure}

\begin{figure}[h]
	\centering
	\includegraphics[width=\largh]{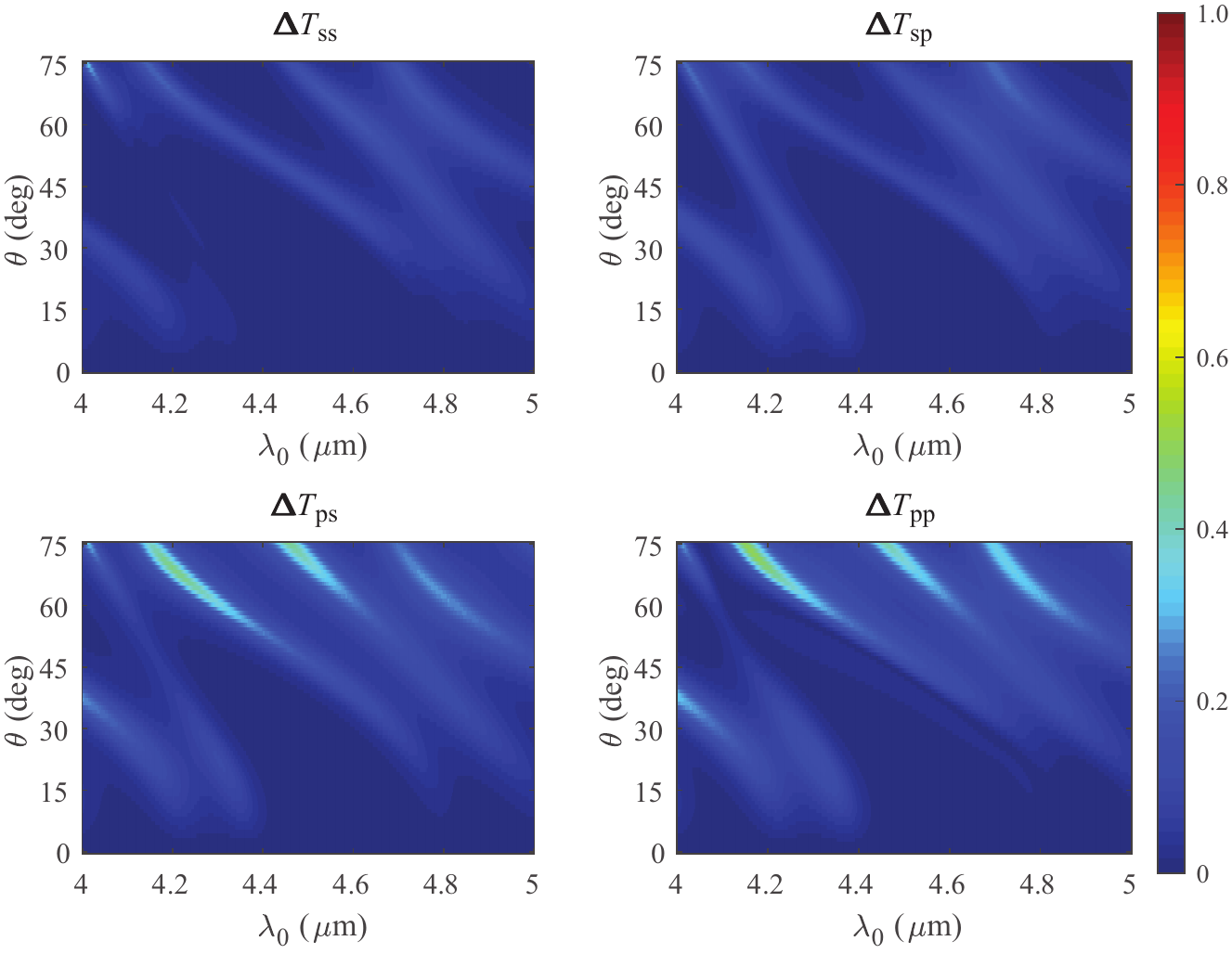}
	\caption{Left/right transmission
		asymmetry functions $\Delta \Tss$, $\Delta \Tsp$, $\Delta \Tpp$, 
		and $\Delta \Tps$ for $\theta\in[0^\circ,90^\circ)$ and $\lambdao\in\quadr{4,5}~\SImum$, when $\psi=45^\circ$ and $N=10$.}
	\label{fig:DeltaT_10}
\end{figure}

When the number of unit cells was increased from $1$ to $10$, left/right asymmetry
appeared for both reflection and transmission. Figures~\ref{fig:DeltaR_10} and 
\ref{fig:DeltaT_10} present density plots of all eight left/right  asymmetry functions
$\Delta R_{\rm ab}$ and $\Delta T_{\rm ab}$, ${\rm a}\in\left\{s,p\right\}$ and ${\rm b}\in\left\{s,p\right\}$, for $\theta\in[0^\circ,90^\circ)$ and $\lambdao\in\quadr{4,5}~\SImum$,
when $\psi=45^\circ$ and $N=10$.  The asymmetry is definitely stronger in transmission than in reflection.

\begin{figure}
	\centering
	\includegraphics[width=\largh]{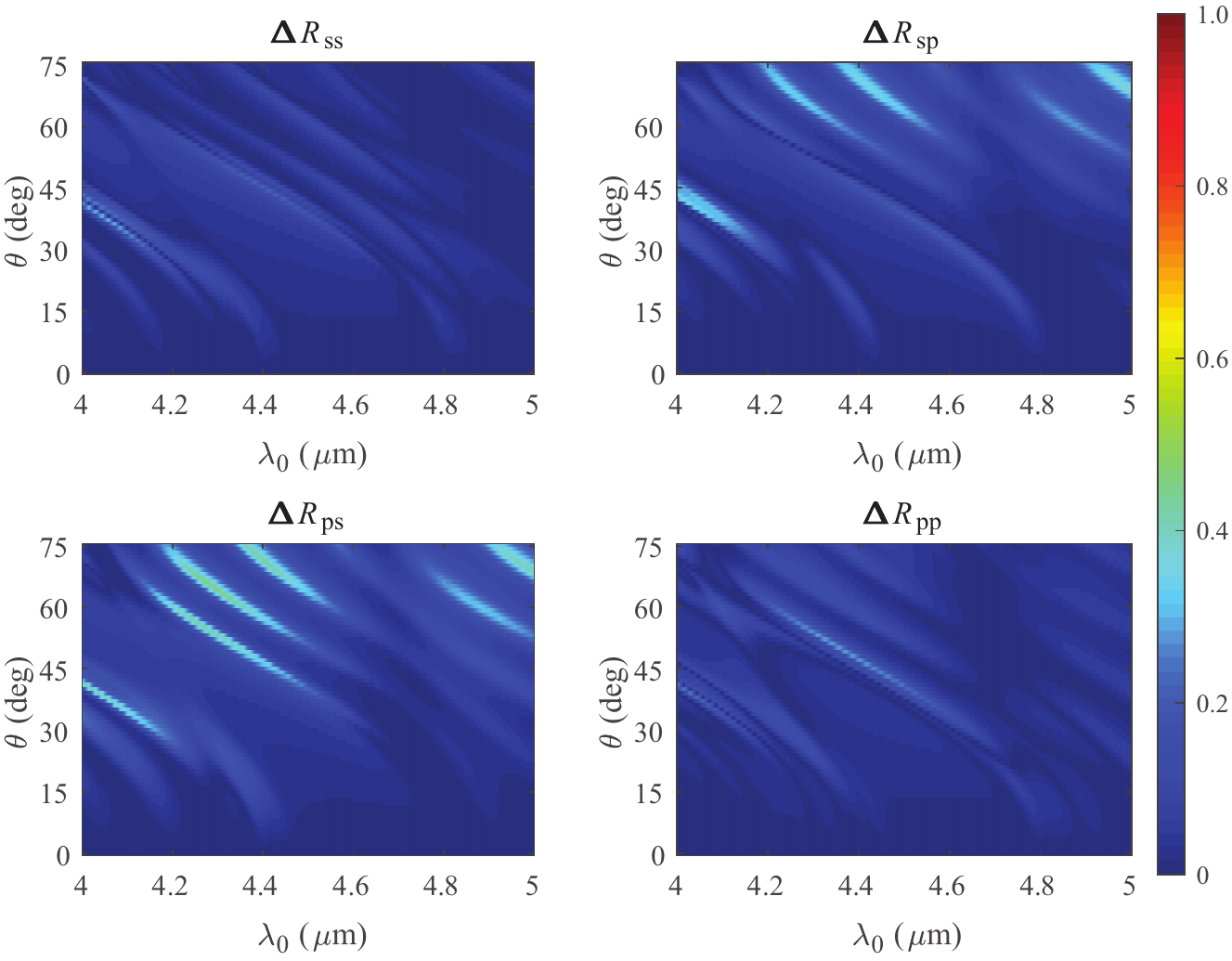}
	\caption{Same as \fref{fig:DeltaR_10} but for $N=20$.}
	\label{fig:DeltaR_20}
\end{figure}

\begin{figure}
	\centering
	\includegraphics[width=\largh]{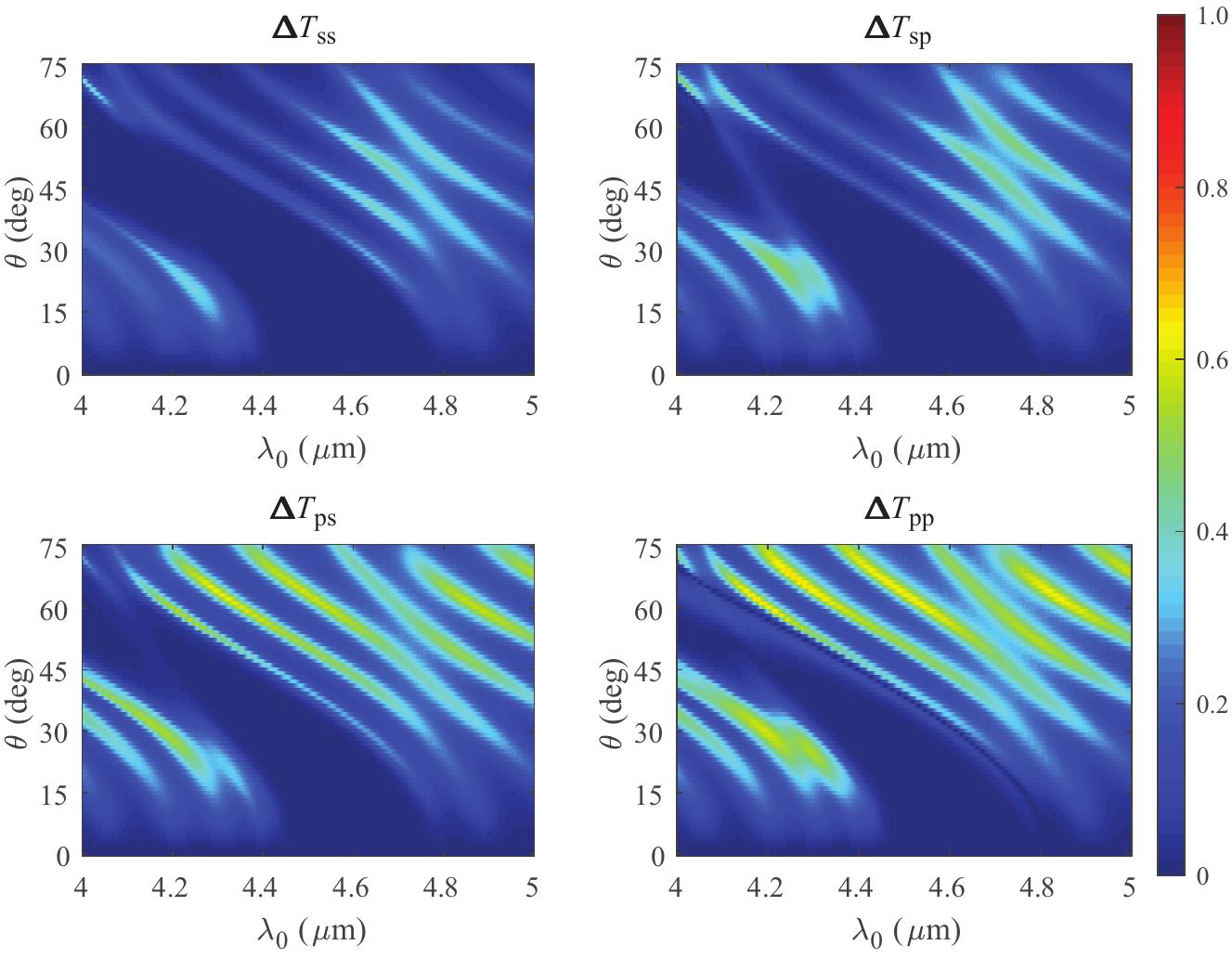}
	\caption{Same as \fref{fig:DeltaT_10} but for $N=20$.}
	\label{fig:DeltaT_20}
\end{figure}

\begin{figure}[h]
	\centering
	\includegraphics[width=\largh]{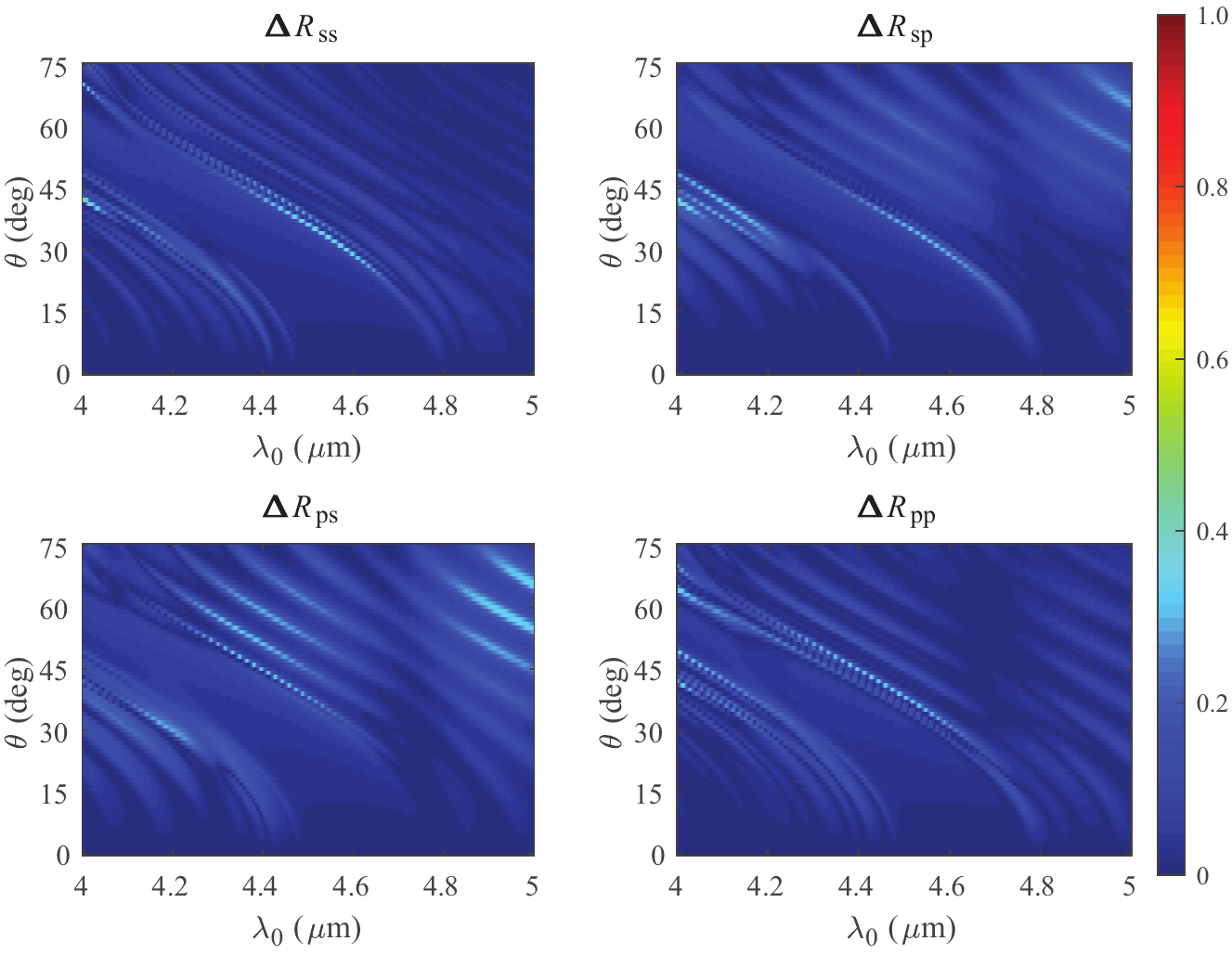}
	\caption{Same as \fref{fig:DeltaR_10} but for $N=30$.}
	\label{fig:DeltaR_30}
\end{figure}

\begin{figure}[h]
	\centering
	\includegraphics[width=\largh]{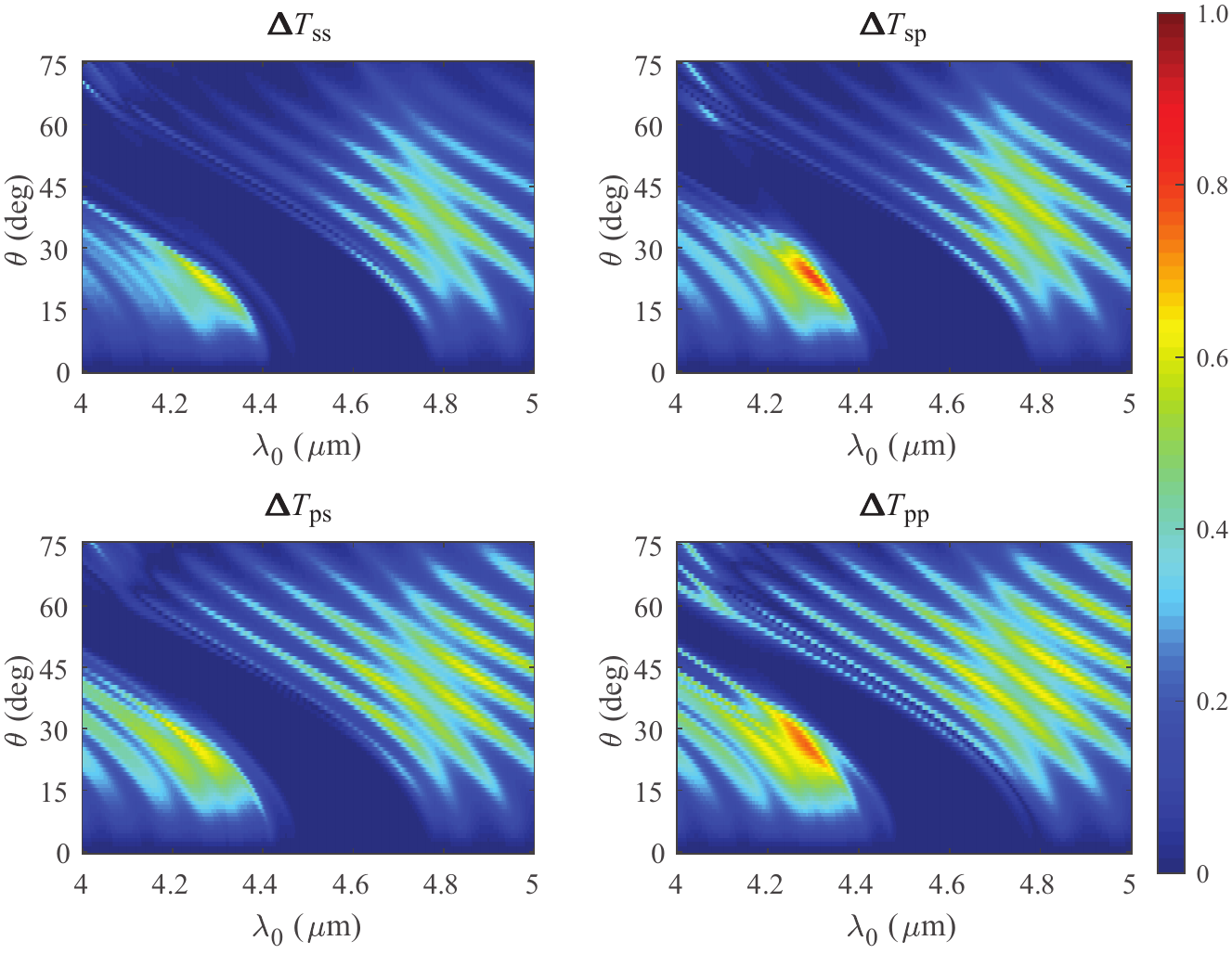}
	\caption{Same as \fref{fig:DeltaT_10} but for $N=30$.}
	\label{fig:DeltaT_30}
\end{figure}

Further increase in the number of unit cells $N$ intensified the left/right asymmetry
in both reflection and transmission, as can be gleaned from
Figs.~\ref{fig:DeltaR_20}
and \ref{fig:DeltaT_20} for $N=20$, and
Figs.~\ref{fig:DeltaR_30}	and \ref{fig:DeltaT_30}
for $N=30$.
Clearly, higher values of the asymmetry functions are obtained with larger values of $N$
and the ranges of $\theta$ and $\lambdao$ are also enhanced thereby.
Additionally, we concluded that:
\begin{itemize}
	\item Left/right asymmetry is stronger for
	the transmittances than for the reflectances.
	\item Left/right asymmetry can be observed for quite wide
	ranges of the incidence angle $\theta$ and the free-space wavelength
	$\lambdao$.
	\item  More left/right asymmetry can be achieved when the angle of incidence is $\theta\gtrsim20^\circ$ while it vanishes, as expected, when $\theta$ approaches $0^\circ$.
\end{itemize}

For additional insights into the density plots of Figs.~\ref{fig:DeltaR_10}--\ref{fig:DeltaT_30}, the maximum values of all eight left/right asymmetry functions were
identified along with the free-space wavelength 
$\lambdao\in\quadr{4,5}~\SImum$ and the incidence angle $\theta\in[0^\circ,90^\circ)$ at which they occur. These data are reported in Tables~\ref{tab:maxDR} and \ref{tab:maxDT} for reflection
and transmission, respectively.  

\begin{table*}[!ht]
		\centering
		\caption{
			{\bf  Maximum values of the left/right reflection asymmetry functions $\Delta \Rab$,
				${\rm a}\in\left\{s,p\right\}$ and ${\rm b}\in\left\{s,p\right\}$,   along with the 
				free-space wavelength $\lambdao$ and the incidence angle $\theta$ at which they occur, when $\psi=45^\circ$.}}
		\begin{tabular}{|c|l|c|c|l|c|c|l|c|c|l|c|c|}
			\hline
			number &\multicolumn{3}{c}{\bf $\Delta \Rss$} & \multicolumn{3}{|c}{\bf $\Delta \Rps$}& \multicolumn{3}{|c}{\bf $\Delta \Rsp$} & \multicolumn{3}{|c|}{\bf $\Delta \Rpp$}\\ 
			\cline{2-13}
			of cells & max. & $\lambdao (\SImum)$ & $\theta (\deg)$ &max. & $\lambdao (\SImum)$ & $\theta (\deg)$ &max. & $\lambdao (\SImum)$ & $\theta (\deg)$ &max. & $\lambdao (\SImum)$ & $\theta (\deg)$ \\
			\hline\hline
			$N=1$ & 0.003 & 4.13 & 75  & 0.009 & 4.09 & 75  & 0.002 & 4.64 & 75  & 0.001 & 4.63 & 75 \\ \hline
			$N=10$ &0.145 & 4.43 & 75 & 0.269 & 4.00 & 36 & 0.256 & 4.18 & 71 & 0.170 & 4.01 & 75\\ \hline
			$N=20$ & 0.294 &  4.05 & 37 &  0.445 &  4.29 & 64 &  0.372 &  4.99 & 68 &  0.278 &  4.36 & 49\\ \hline
			$N=30$ &  0.463 &  4.01 & 41 &  0.449 &  4.99 & 65 &  0.437 &  4.00 & 41 &  0.380 &  4.01 & 41\\ \hline
		\end{tabular}
		\label{tab:maxDR}
\end{table*}

\begin{table*}[!ht]
		\centering
		\caption{{\bf  Maximum values of the left/right transmission asymmetry functions $\Delta \Tab$,
				${\rm a}\in\left\{s,p\right\}$ and ${\rm b}\in\left\{s,p\right\}$,
				along with the free-space wavelength $\lambdao$ and the incidence angle $\theta$ at which they occur, when $\psi=45^\circ$.}}
		\begin{tabular}{|c|l|c|c|l|c|c|l|c|c|l|c|c|}
			\hline
			number &\multicolumn{3}{c}{\bf $\Delta \Tss$} & \multicolumn{3}{|c}{\bf $\Delta \Tps$}& \multicolumn{3}{|c}{\bf $\Delta \Tsp$} & \multicolumn{3}{|c|}{\bf $\Delta \Tpp$}\\ 
			\cline{2-13}
			of cells & max. & $\lambdao (\SImum)$ & $\theta (\deg)$ &max. & $\lambdao (\SImum)$ & $\theta (\deg)$ &max. & $\lambdao (\SImum)$ & $\theta (\deg)$ &max. & $\lambdao (\SImum)$ & $\theta (\deg)$ \\
			\hline\hline
			$N=1$  & 0.005 & 4.09 & 75  & 0.006 & 4.08 & 75  & 0.005 & 4.08 & 75  & 0.006 & 4.08 & 75  \\ \hline
			$N=10$ & 0.340 & 4.01 & 75 & 0.454 & 4.20 & 68 & 0.235 & 4.03 & 72 & 0.497 & 4.17 & 72 \\ \hline
			$N=20$  &  0.414 &  4.66 & 51 &  0.617 &  4.32 & 62 &  0.491 &  4.23 & 25 &  0.654 &  4.27 & 66 \\ \hline
			$N=30$  &  0.629 &  4.27 & 22 &  0.654 &  4.24 & 26 &  0.796 &  4.31 & 22 &  0.752 &  4.29 & 25 \\ \hline
		\end{tabular}
		\label{tab:maxDT}
\end{table*}

The first rows of both    Tables \ref{tab:maxDR}, and \ref{tab:maxDT} confirm that  left/right asymmetry in reflection as well as transmission exists for $N=1$, but is extremely weak. Our multilayering strategy, however, is successful in that maximum values of all eight left/right asymmetry functions increase as
the number of unit cells---which is the same as the number of TI layers---increases. However, none of the eight asymmetry functions can increase indefinitely and substantially with increasing $N$, 
because they are bounded as follows:
\begin{equation}
\left.\begin{array}{l}
0\leq\Delta\Rab\leq1
\\[5pt]
0\leq\Delta\Tab\leq1
\end{array}\right\}\,,
\quad{\rm a}\in\left\{\rm s,p\right\}\,, \quad {\rm b}\in\left\{\rm s,p\right\}\,.
\end{equation}
In other words, there will be diminishing returns for $N$ exceeding some $\overline{N}$
for any specific value of $\overline{\gamma}$.

\section{Concluding Remarks}\label{sec:cr}
In an attempt to achieve a high left/right asymmetry using a topological insulator  with a feasible value of the surface admittance $\gammaTI$ to quantitate protected surface states, we proposed and investigated a periodic-multilayer structure made of a topological insulator alternating with an anisotropic material with columnar morphology. 

Analysis was performed by varying both the free-space wavelength and the direction  of incidence,
Left/right asymmetry is definitely evinced in both reflection and transmission
by a single TI layer partnered with a layer of an anisotropic dielectric material, but the asymmetry is so weak as to be technologically unattractive. Given that TIs with larger values of $\gammaTI$ are presently unavailable and that the magnetic routes can probably just double or treble the magnitude of the surface admittance, the multilayering strategy proposed here offers a way to enhance left/right asymmetry in a major way, both in reflection and transmission. There will be, for sure, some maximum number of
unit cells in the periodic multilayer beyond which  increases in left/right asymmetry will greatly diminish,
because the asymmetry functions are bounded. Still, high degrees of left/right asymmetry are going to be available, because both TI layers \cite{DiPietro} and dense columnar thin films \cite{HWbook}
can be deposited using standard physical-vapor-deposition techniques \cite{Mattox,Macleod,Raul,Baumeister}.

\vspace{5mm}

\noindent {\bf Acknowledgment.}
  AL thanks the Charles Godfrey Binder Endowment at Penn State for ongoing support of his research.

\end{document}